\documentclass[conference]{IEEEtran}[10pt]
\usepackage{amsmath}
\usepackage{amssymb}
\usepackage{graphicx}
\usepackage{subfigure}
\usepackage{a4wide}
\usepackage{algorithm}
\usepackage{algorithmic}
\usepackage{amsfonts}

\begin{document}
\title{Data-centric Misbehavior Detection in VANETs}
\author{Sushmita Ruj$^*$, Marcos Antonio Cavenaghi$^\dag$, Zhen Huang$^\ddag$, Amiya Nayak$^*$, and Ivan Stojmenovic$^*$\\ 
$^*$SITE, University of Ottawa, Canada -- \{sruj,anayak,ivan\}@site.uottawa.ca\\
$^\ddag$SITE, University of Ottawa, Canada -- zhuan045@uottawa.ca\\
$^\dag$Unesp, Sao Paulo State University, DCo, Brazil -- marcos@fc.unesp.br}
\maketitle

\newcommand{\ot}{\otimes}
\newcommand{\m}{\mu}
\renewcommand{\l}{\lambda}
\newtheorem{definition}{Definition}
\newtheorem{construction}{Construction}
\newtheorem{theorem}{Theorem}
\newtheorem{question}{Question}
\newtheorem{lemma}{Lemma}
\newtheorem{proposition}{Proposition}
\newtheorem{remark}{Remark}
\newtheorem{corollary}{Corollary}
\newtheorem{example}{Example}

\begin{abstract}
Detecting misbehavior (such as transmissions of false information) in vehicular ad hoc networks (VANETs) is very important problem with wide range of implications including safety related and congestion avoidance applications.
We discuss several limitations of existing misbehavior detection schemes (MDS) designed for VANETs.
Most MDS are concerned with detection of malicious nodes. 
In most situations, vehicles would send wrong information because of selfish reasons of their owners, e.g. for
gaining access to a particular lane.
Because of this (\emph{rational behavior}), it is more important to detect false information than to identify misbehaving nodes. 
We introduce the concept of data-centric misbehavior detection and 
propose algorithms which detect false alert messages and
misbehaving nodes by observing their actions after sending out the alert messages.
With the data-centric MDS, each node can independently decide whether an  information received is correct or false. The decision is based on the consistency of recent messages and new alert with reported and estimated vehicle positions.
No voting or majority decisions is needed, making our MDS resilient to Sybil attacks.
Instead of revoking all the secret credentials of misbehaving nodes, as done in most schemes,
we impose fines on misbehaving nodes (administered by the certification authority), discouraging them to act selfishly.
This reduces the computation and communication costs involved in revoking all the secret credentials of misbehaving nodes.

\end{abstract}


\section{Introduction}
Vehicular ad hoc network (VANET) consists of vehicles (also referred to as nodes), 
road side units (RSUs) and certification authorities (CAs), whose
goal is to ensure road safety and help in secure transfer of message and data. 
Communication can either be vehicle-to-vehicle (V2V) (e.g. relaying alert information)
or vehicle-to-infrastructure (V2I) (e.g. when the vehicle
needs to report some event to the CA). 
Security in VANETs is important, because the message sent by one vehicle might have important consequences such as accident prevention. 

VANETs are a class of ephemeral networks ~\cite{R09}, where the connection between vehicles (nodes) is very short lived. 
The network topology changes very frequently, as nodes move in and out of range of each other. 
The density of the network also changes over time, e.g. during rush hours. 
These characteristics make VANET very challenging for dealing with security issues. 

Each vehicle has an on board unit (OBU), which broadcasts messages about the position, speed, acceleration/deceleration, 
alert signals etc. OBU also has authentication capabilities, to
verify that an incoming message has been broadcasted by a valid entity. 
Roadside units (RSUs) help in coordinating vehicle activities and 
collect information about nearby vehicles and their actions (e.g. red light violations). Human behavioral tendencies will be reflected in the movement of the vehicles (rational behavior).
Vehicles (faulty nodes) can either start malfunctioning due to some internal failures and give out false alerts,
false location and speed intentionally for selfish reasons. 
Malicious vehicles may also attempt to gather sensitive information about other users e.g. credit card number while
interpreting RFID signals 
at an electronic toll station. 

To protect its privacy, each node does not use its unique identity (for example, the electronic license plate), 
but \emph{pseudonyms}, when broadcasting data. 
Pseudonyms are generated and assigned in such a way that a node's unique identity cannot be derived by observing two or more pseudonyms. 
Users can also authenticate themselves using pseudonyms. 

Current research on security in VANETs has been focussed on location privacy, maintaining authenticity of data and revocation of certificates and
secret credentials.
Surveys on the security challenges in VANETs can be found in \cite{PBHSFRMKKH08,KPBMSWTCHKH08,PP05,BHM10}. 
Most papers on location privacy deal with how  to assign pseudonyms \cite{R09}, when to change pseudonyms \cite{R09,SLHP07,FMBH10,BHV07}, and
how to assign signatures using pseudonyms \cite{SLLSS10}. 
Authentication techniques rely on signatures, such that a message is signed with a private key which can be verified 
if a user has the corresponding public key.
A certificate is also issued which verifies the validity of the public key.  
Signature schemes for VANETs have been studied extensively, e.g. ECMV \cite{WJS08} and PASS \cite{SLLSS10}. 
Revocation of malicious nodes is another issue that has received a lot of attention. 
The issues are whether to maintain a list of all revoked certificates and keys or revoked vehicles or some seed of the revoked vehicles \cite{SLLSS10}.   
Revocation of certificates and secret credentials has the following disadvantages.
The certificate revocation list (CRL) containing all the certificates of revoked vehicles,  
has to be sent to all the nodes in the network.
This approach requires a huge bandwidth, if the number of revoked nodes is high. 
Next, revocation may not be necessary if a vehicle misbehaved only once for some selfish reason.

In this paper we assume that nodes misbehave mostly because of selfish reasons, to reach their destinations faster.
For example, vehicle might send false report on congestion, accident or road block. 
It is conceivable to believe that a vehicle does not have malicious intentions of causing accidents.
Each vehicle normally sends valid and useful information.
If all the certificates are revoked then useful information sent will be ignored. 
Therefore we argue that we do not need to classify vehicles according to their overall behavior, but instead to distinguish between correct and false information received from a vehicle. 
For example, how to verify the report about an approaching vehicle (possible emergency vehicle)? 
Therefore, it is important to identify false data and the sender efficiently, because a delay of even one second might cause traffic accident. 
This problem is termed as \emph{data-centric misbehavior detection} in contrast to entity-centric misbehavior detection, where the main goal is to find out and penalize a misbehaving node.
The idea of data-centric misbehavior detection stems from Raya's work \cite{R09} on data-centric trust, where the author considers trust on information
rather than on the source of information. 
In our approach, we do not revoke nodes which misbehave. Instead,
the misbehaving node receives a fine, depending upon its action. 
It can keep on sending information which might not necessarily be malicious. 
The payment of fines would hopefully discourage nodes from sending further false messages.

We will concentrate on detecting false alert messages and false location information sent out by a node. 
We would like to detect alerts like  emergency breaking, approaching emergency vehicles, road feature notifications, change of lanes etc. 
A list of such alerts is given in Section \ref{subsec:model}.

Intrusion detection has been studied extensively in the context of wireless ad hoc networks. However the existing solution approaches are not applicable for detecting 
malicious behavior in VANETs. 
In Section \ref{subsec:IDS} we will discuss intrusion detection schemes
used for other ad hoc networks and explain why they cannot be applied to VANETs. In a nutshell, the detection itself is application and scenario dependent. Trust management based solutions are not feasible here because neighborhood may change rapidly and therefore trust relationships could be short lived and difficult to even establish in the first place. In relatively static neighborhood graphs (e.g. in congested areas) neighbors may not have history of misbehavior so the first violation cannot be automatically detected. Central authority may not be available to facilitate misbehavior detection and penalize accordingly.

The first existing solution to misbehavior detection problem in VANETs, by Golle et al \cite{GGS04}, creates a model of the network. 
The model of the network is the set of all possible events in the network. 
An event which is observed by another node is checked with the model. If it is valid, according to the model, 
then it is considered to be a correct message, otherwise false. 
The main problem with this approach is that it has not been shown how this model can be created and maintained. 
For VANETs, which consists of several nodes, building up a global database (as pointed out in \cite{GGS04})
can be very expensive and impractical. 
Another problem is that the scheme does not provide location privacy, which our scheme achieves using pseudonyms.

Some solutions \cite{ZCNC07,PATZ09} are based on countering Sybil attacks \cite{D02} in VANETs. 
The schemes assume that the precise location of nodes are known. 
Moreover, these scheme cannot detect false alerts raised by nodes. 
In Sybil attacks nodes pose as separate identities and influence the decision of revocation and MDS, which rely on majority votes.
Since our scheme takes into account individual decisions, Sybil attacks poses no threat to the network.

Ghosh et al \cite{GVGKM10} investigated post crash scenarios. 
They compared the expected and actual trajectory to decide if a node is sending the correct post crash notification (PCN) alert. 
The expected trajectory has been modelled using node's possible behavior. 
For example a lazy node might not take any action until it is very close to the site of crash.
On the other hand a risk-averse node might move away very far from the site of crash.  
There are three aspects to be noted: the modelling of expected trajectory, the reported position of the node and the actual position of the node. 
There are two major drawbacks in their scheme.  
Firstly, they assume that a malicious node always sends its correct location information. 
This is not a valid assumption, because the nodes might send wrong location information and compel other nodes to believe
that their trajectory is what is expected. 
Even a small change in position can make a huge difference, for example lane change. 
Secondly, the actual trajectory may indeed legitimately differ from the one predicted by modeling the movement. For example a car might turn right at a crossing or prepare for left turn
and change the lane. These have not been considered by Ghosh et al \cite{GVGKM10}. 

\begin{figure}[htb]
        \centering
        \subfigure[Node $n_j$ sends alert ``Road block at location X"]{

            \includegraphics[width=2.5in]{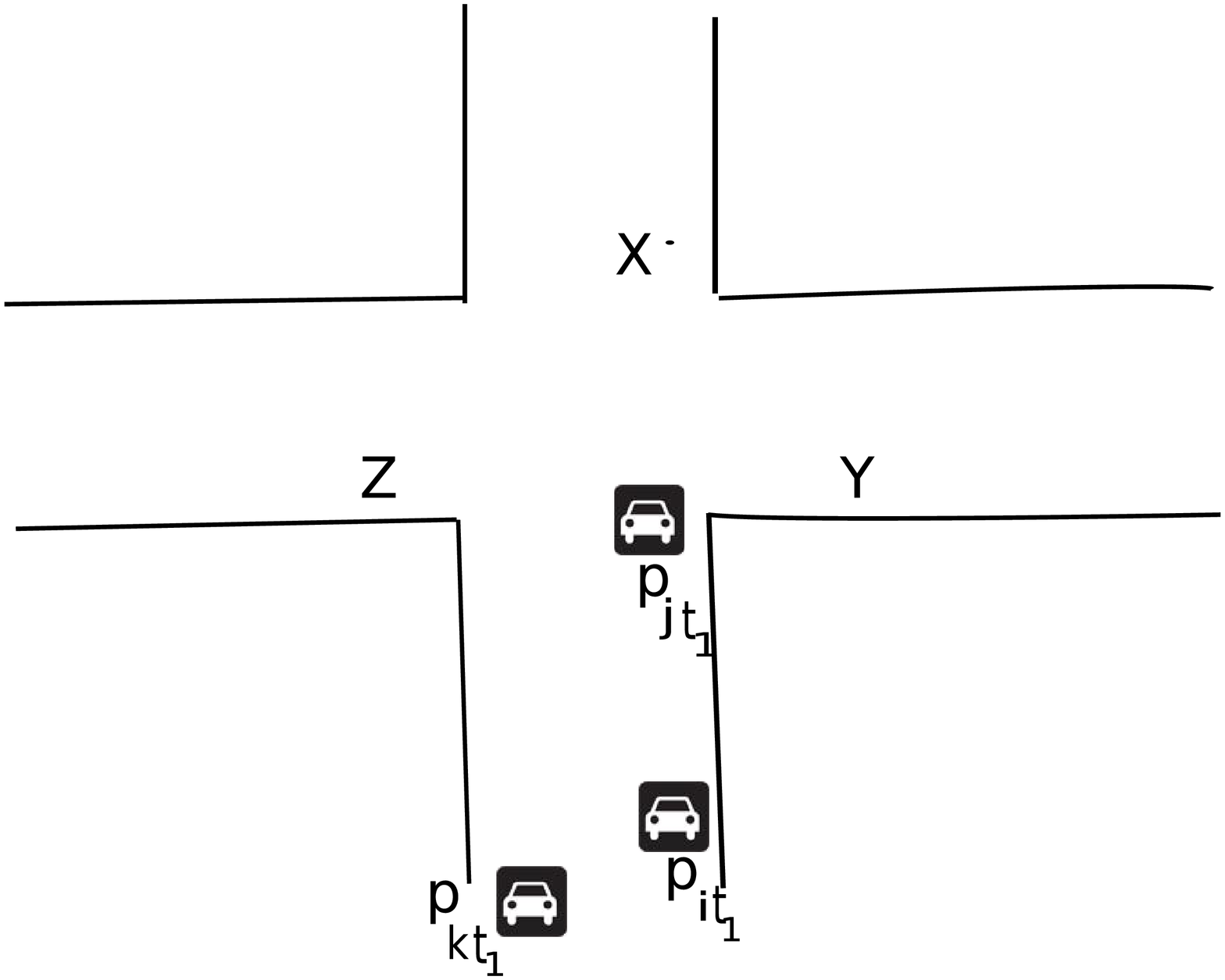}
            \label{fig:subfig1}

        }
        \subfigure[Node $n_j$ goes past the location $X$]{

            \includegraphics[width=2.5in]{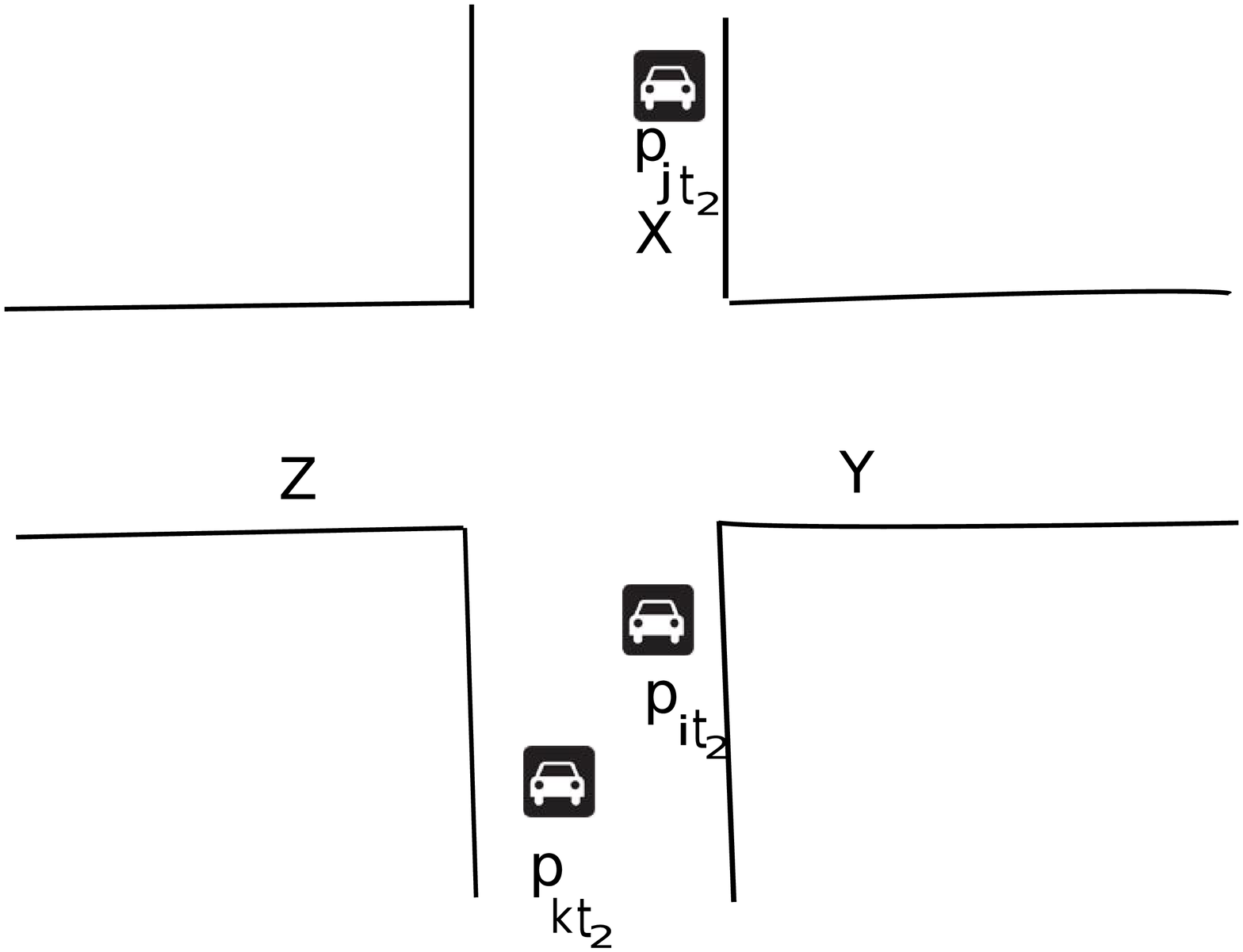}
            \label{fig:subfig2}

        }
        \label{fig:congestion}
        \caption{Example: inconsistencies in messages prove misbehavior} 

    \end{figure}

In this paper we propose techniques to detect misbehaving nodes and incorrect data, which will also preserve the privacy of the network. Vehicles sent periodic beacon messages, so that the positions of neighbors is monitored over time. Our method borrows some ideas form \cite{GVGKM10}.
If reported position is not consistent with the alert raised then the receiving node declares the message as incorrect and discards it. 
Consider the situation in Fig \ref{fig:congestion}. 
At time $t_1$ node $n_j$ (with pseudonym $p_{jt_1}$) sends an alert that ``Road block at location X" (Fig \ref{fig:subfig1}) 
At time $t_2$ ($t_1$ close to $t_2$)  its position is past $X$ (Fig \ref{fig:subfig1}).  
This suggests that either there is no road block  at location  X, or location information of $n_j$ is wrong. 
It might do so, to divert the traffic to another lane, to gain easy access of lane X. 

We address more general types of reported information compared to \cite{GVGKM10}. Nodes can send any kind of information, either different alerts or lane changing information.
The other cars should be able to verify the validity of the message content from the subsequent location information.
We also address the problem of verifying  the validity of location information reported by neighbors. 
If wrong information is detected, the CA is contacted via the nearest RSU.
CA knows the mapping of the pseudonym to the actual node ID of the misbehaving node. 

Convicted vehicle is not revoked. Instead, CA imposes a fine on it.  
Since the nodes change their pseudonyms on a regular basis, it might not be possible
to link the alert message sent and the location information as coming from the same vehicle. 
We therefore impose the restriction on the lifetime of pseudonyms. There should be a certain time interval
after sending an alert message, before the pseudonym can be changed. 

\subsection{Our contribution}
\begin{enumerate}
\item We propose a new model of VANET where we assume that most misbehaviors arise out of selfish motives. However, 
our model can also handle misbehavior from malicious nodes. 
\item We do not revoke misbehaving nodes, but impose fines on them. 
This reduces the communication and computation costs in calculating,  transmitting, and storing  certificate revocation lists. 
\item Misbehavior is detected by observing alerts raised by a node and its subsequent action. 
\item Our approach does not rely on voting schemes and group associations. Therefore it is immune to Sybil attacks.
\item False location information can be detected in addition to detecting false alert messages. 
\end{enumerate} 

\subsection{Organization}
The paper is organized as follows. In Section \ref{sec:related} we present related work on VANET security. 
We discuss the limitations of existing MDS in VANETs in Section \ref{sec:existing_MDS}. 
Our network model, definitions and notations are presented in Section \ref{sec:model_def_not}. 
We present our misbehavior detection scheme in Section \ref{sec:MDS}.
We discuss the limitations of our scheme in Section \ref{sec:limitations} and conclude in Section \ref{sec:conclusion}. 

\section{Related work}
\label{sec:related}
In this section we discuss security and privacy issues in VANETs. 
We then discuss existing intrusion detection schemes in general ad hoc networks and explain why the are not a good option for misbehavior detection 
in VANETs.

\subsection{Security and privacy in VANETs}
\label{subsec:sec_priv}
Security and privacy  in VANETs involve the following important issues: authentication, location privacy, misbehavior detection and 
revocation. Amongst these authentication, location privacy and revocation has received a lot of attention. 
In this section we give a brief overview of the existing work on these three issues.

Authentication is done using two techniques: 
1) group signature schemes and 2) pseudonyms.
Pseudonyms also help in privacy protection. 
In a signature scheme, each user is given a private key, with which it signs the message.

Each user can construct  the public keys of all the users. 
When another user receives the signed message, it can verify the signature and check the authenticity of the message. 
Group signatures were introduced by Chaum and van Heyst \cite{CH91} 
to provide anonymity to the signers.

Boneh et al \cite{BBS04} suggested the use of group signatures in vehicular networks. 
Group signatures \cite{SLHP07,CPHL07,SSBP09} can be applied to sign message in VANETs, so that, when  another vehicle receives
the message, it  can only check the authenticity of the message, 
with no means to track the node  who sent it. 
Although these schemes provide authentication, conditional anonymity and non-repudiation, they result in large
revocation costs. 
Since groups can change very frequently in a city network, this scheme is not so practical. 

The existence of a single identity poses a great threat to the privacy of a user.
A user's movement can be easily tracked. It is also possible to trace the  personal details of the user. 
For this reason each vehicle is given a set of aliases which are called \emph{pseudonyms} and a unique identity. 
The unique identity is known only to the user and a certified authority which issues the pseudonyms.
Any other node or RSU only knows the pseudonym. 

The pseudonyms are generated in a way that the identity of the node cannot be obtained from the pseudonyms. 
A vehicle can also have multiple public/private key pairs, corresponding to each pseudonym. 
This concept of \\pseudonyms was introduced by Hubaux et al \cite{HCL04} and has gained a lot of attention.  
Pseudonyms were used in authentication in \cite{RH07}.
Calandriello et al \cite{CPHL07} used a hybrid scheme using pseudonyms and group signatures for authentication.

Pseudonyms are changed from time to time to preserve location privacy. 
In \cite{FRFPH07}, the authors point out that the pseudonyms should be changed only in the \emph{mix zones}.
Mix zones are areas where nodes  cannot be observed, either
by another node or by a RSU. 
The problem with this approach is that if there is only one node
in a mix-zone and it changes its pseudonym, then it is clear that the two pseudonyms belong to the same node. 
However, if there are more than one vehicle in the mix zone and they change their pseudonym, 
then it cannot be easily predicted which pseudonym corresponds to which node. 
Buttyan et al \cite{BHV07} show by simulation, how the privacy level is changed using the above approach. 
Frequent change of pseudonyms ensure higher privacy, but pseudonyms are expensive and often obtained from
a central authority. 

Freudiger et al \cite{FMBH10} give a detailed study about the age of pseudonyms and discussed different parameters, on which
the age of pseudonym depends. 
Sampigethaya et al \cite{SHLPM05,SLHP07} used random silent period between the change of pseudonyms. 
They assumed that vehicles move in a group with similar speed. When a new vehicle joins the network with some pseudonym, 
it waits for a random time before changing its pseudonym. 
For example, if two nodes enter the network
at the same time and change their pseudonym after a random time interval, then the new and old pseudonyms of a vehicle cannot be linked. 

Another area of vehicular network security which has received a lot of attention is revocation of nodes. 
Most of the work in this area assumes that there is an underlying misbehavior detection mechanism, 
that has detected the misbehaving nodes. 
Revocation can either be local or global.
 
In \cite{RPAJH07}, the authors present a \emph{local revocation} scheme using \emph{LEAVE} protocol. 
The misbehaving node is revoked from the neighborhood, generally by voting. 
Moore et al \cite{MRCPAH08} used a ``suicide" mechanism called \emph{sting}, in which revocation is done locally. 
A node accusing a misbehaving node is also blacklisted by the neighboring nodes, along with the accused node. 

This sacrificing behavior demonstrates that the first node is honest. 
This scheme can be attacked in the following way:
Consider the situation in which there is a benign node surrounded by misbehaving nodes.  
Once honest node issues accusation signal, it is revoked and
cannot accuse the other misbehaving nodes. 
Also, if one of the misbehaving nodes accuses a honest node and is revoked, the other malicious nodes still remain in the network. 
A misbehaving node might accuse  another misbehaving node and be considered honest by other nodes.

Two game theoretic based revocation schemes \cite{RMFH08,BMRH10} have been proposed. 
Nodes can either vote, abstain for voting or commit ``suicide". 
Each of these actions have associated pay-off and costs. Nodes choose the action that maximizes their benefits. 
A recent work by Liu et al \cite{LCH10} show the limitations on revocation in VANETs using the game theoretic approach. 
The above papers assume that the number of neighbors is known which is not the case for VANETs. 
The CA decides the costs and pay-offs and might not be available all the time. 

Revocation of nodes means revocation of their certificates. 
However, the use of pseudonyms imply that the malicious node's certificates must all be revoked. 
This increases the size of the certification revocation list (CRL). In \cite{LHH08,PMH08}, the CRL is
transmitted from vehicle to vehicle and is therefore require significant communication overhead. 
Recently, 
Sun et al \cite{SLLSS10} proposed an  authentication scheme in which the CRL size is of the
order of the revoked vehicles and does not depend on the number of pseudonyms that the revoked vehicles have.   

\subsection{Intrusion detection schemes in ad hoc networks}
\label{subsec:IDS}
There has been several intrusion detection schemes for ad hoc networks. A survey can be found in \cite{AW08}. 
In some schemes like SCAN \cite{RBK05,YSML06} neighboring nodes monitor a given node. 
If the number of votes against the node exceeds a threshold, then the node is evicted. 
This will not work for VANETs because of the ephemeral nature of the network, 
there might not be enough nodes in the vicinity of a malicious node
to revoke it. Alert based schemes \cite{GS04,ZYN08} raise an alert either to the base station or other nodes, 
if abnormal behavior is observed.  
Malicious nodes might raise false alerts against benign nodes and jeopardize the situation. 
Base stations might not be within communication range to analyze the observed behavior. 
If there are too many malicious nodes in the vicinity of a benign node, they might falsely accuse it, but collectively raising alerts against it. 
Clulow and Moore's \cite{CM06} uses a suicide scheme where a node which convicts another node has to sacrifice one of its certificates. 
The whole idea of suicide will deter a malicious node to accuse a benign node. 
They use a game theoretic approach. 
Reidt et al \cite{RSB09} adds the concept of \emph{karmic-suicide} which gives incentives to nodes which commit suicide and
includes them in the network again. 
Game theoretic techniques cannot be used as is, in misbehavior detection in VANETs, because in an ephemeral network, 
costs and benefits will change over time and vary from node to node. 
Two questions that need to be answered are who decides these costs and benefits and how these costs and benefits change in the network.

\section{Limitations of existing misbehavior detection schemes in VANETs}
\label{sec:existing_MDS}
In this section, we discuss various MDS in VANETs. We point out their limitations, which motivated us
to design a new misbehavior detection scheme. 

The first paper on misbehavior detection in VANETs was by Golle et al \cite{GGS04}. 
The paper proposes an approach to detect malicious data based on its deviation from an existing  model of the network. 
They adopt a parsimony argument which assumes that an attack involving a few malicious node is 
more likely to happen than a collusion attack consisting of a large number of nodes. 
Each event has an associated location information. 
Nodes observe different events and store them in a global database. 
The \emph{model of the VANET} lists the set of all possible events. 

First, maintaining a global database of events  might not be easy for VANETs because of the large network size. 
A global database cannot be maintained by the nodes and maintaining it with
a trusted authority might make it time consuming to compare the results with the observed values. 
Even if only the local information is maintained in the database, it has to be changed from time to time
as nodes enter and leave the neighborhood. 
Maintaining such a database means that entries have to be added and deleted frequently which is not effective. 

Second, Golle's scheme cannot provide privacy by change of pseudonyms. 
If pseudonyms are allowed, the size of the database increases drastically, because each node can have several representations.
Also, there might not be enough data gathered for a node having a certain pseudonym. 

Sybil attack \cite{D02} pose a great threat to VANETs.
In a Sybil attack, a malicious node creates several false identities and poses as multiple vehicles. 
False information reported by such a node will be convincing to the rest of the network, because
it appears that several entities agree with the information. 
This can be very damaging where a false information might lead to accidents. 
Since currently most malicious detection algorithms and revocation schemes use majority or voting, 
Sybil attack becomes a serious problem in VANETs. 

There has been several papers to counteract Sybil attacks  in VANETs. 
In \cite{ZCNC07}, the authors propose a scheme called Privacy-preserving Detection of Abuses of Pseudonym P$^2$DAP. 
In P$^2$DAP, there is a large pool of pseudonyms. 
There are two hash functions $H_c$ and $H_f$, which are  called coarse grained and fine grained hash functions respectively. 
Each pseudonym has a fine grained hash value $H_f(p_i|k_f)$ and a coarse grained hash value $H_c(p_i|k_c)$.
($p_i$ is the pseudonym and $k_f$ and $k_c$ are keys which are used from freshness). 
The fine grained hash value of all the pseudonyms of a node are the same. 
Any two fine grained hash value of pseudonym belonging to different nodes is different. 
When a RSU (referred in the paper as road-side boxes) receives a message, it notes the pseudonym that was used to
sign the message. 
If there are two pseudonyms that hash to the same course grained value, then either they are from the same vehicle (which
has several entities) or from different vehicles which hash to the same coarse grained value. 
The RSU sends these values to the CA, which checks the fine grained values to see if the same message
was send by a node with multiple entities or from several nodes.  
In this way Sybil attack is detected. 
If more than one set of coarse grained values are present, then all these are send to the CA, which verifies if there is a Sybil attack. 
The scheme can be used to detect collusion attacks of size less than some threshold, $\tau$. 
Suppose there are $|S_c| < \tau$ coarse grained values, 
then the RSU sends the pseudonyms of the corresponding hash values to the CA, which checks if they are from the same nodes. 

The scheme assumes that the misbehaving vehicles are within the reach of some RSU, which is not a 
valid assumption. Since the scheme relies on collecting evidence from several nodes and calculating all hash values, 
the detection can take as much as 200 seconds (as stated in the simulation results). 
This might be too unrealistic in practical situation where prompt action might be needed to prevent a crash. 
No action is taken against compromised RSUs. 

Xiao et al \cite{XYG06} which use signal strength to detect Sybil attacks, suffers from the defect that a Sybil node might use
different signal strength to send different messages and will not be detected. 
In \cite{PATZ09}, the RSU assigns certified time-stamps to all vehicles that pass by it. 
When a message is sent out by a node, a series of recent time stamp certificates are also send out. 
It can be assumed that  two or more nodes cannot  pass the same RSU at the same time for several times. 
So, if the same series of time stamps are observed for two or more nodes, then it can be assumed
that the node is a Sybil node. 
A malicious node might change the time stamps
and send different time stamps for different identities that it has faked, and avoid being detected as a Sybil node. 

Raya et al \cite{RPAJH07} proposed a scheme to detect and revoke malicious nodes. 
Each node has several pseudonyms. 
Corresponding to each pseudonym, there is one public/private key pair and a certificate issued by the CA. 
To revoke a node, its certificates are revoked. 
Their scheme consists of three components:
(1) Revocation of Trusted Component (RTC), 
(2) a Misbehavior Detection System (MDS), and
(3) a Local Revocation Protocol by Voting Evaluators(LEAVE).
The MDS system observes the behavior of a node and compares it with the average behavior. 
This is done using entropy. If $p_i$ is the probability that node $i$ is an attacker, then the entropy
$H = \sum_{i=1}^N p_i \log p_i$, (where $N$ is the number of nodes in the network) will be low if few nodes behave differently and will be high
if many nodes behave differently. 
The $K$-means clustering algorithm \cite{JMF99} is then used to find out the exact misbehaving node. 
Is there are many malicious nodes, then the benign nodes will be convicted. 
The LEAVE protocol evicts the certificates of a node, if the number of accusations is above a certain threshold. 
Since the nodes change their pseudonyms, there might not be enough evidence against a malicious node if there are too few nodes in its neighborhood. 
LEAVE requires an \emph{honest majority}, meaning that a benign node must be surrounded by more benign nodes than selfish/malicious ones. 
So, the scheme fails when there are too many misbehaving nodes around a benign node.

Moore et al \cite{MRCPAH08} propose a scheme Stinger for faster exclusion of misbehaving nodes. 
Stinger is faster than LEAVE, however, LEAVE has a lower false positive rate. This means that
it revokes fewer benign nodes than Stinger. 
Stinger works as follows: when a benign node $G$ detects a bad node $B$, then it broadcasts a $sting_{G,B}$. 
All node near $G$ blacklists both the nodes $B$ and $G$. Any message from $B$ and $G$ is disregarded. 
However, they can still receive and forward messages. 
When the nodes are within the reach of RSU, they are evicted altogether. 
The disadvantage with Stinger is that 
many benign nodes can be ignored to evict one bad node, as has already been pointed out by the authors. 
If several bad nodes are present in the vicinity of a benign node, then this method becomes ineffective. 

A recent paper by Ghosh et al \cite{GVGKM10} detects the root-cause of a misbehavior, in order to determine 
the future action to be taken. 
They address post-crash notifications (PCN), where nodes send false information about a PCN alert, after an accident has taken place.  
The malicious node might either send a crash alert, even if there is no crash and not send a crash alert, 
even if there is a crash.
This situation has been  considered in \cite{GVKG09}.  
The analysis is based on the deviation of the actual trajectory from the expected trajectory. 
For this reason, the trajectory of the node sending out the alert is divided into predefined sample points. 
The node's location is sensed at these sample points.
The expected trajectory is calculated and compared against the sensed trajectory. 
Depending upon the deviation from the expected values, the type of misbehavior can be detected. 
There are several issues to consider in this approach. 

The change of pseudonyms during the estimation of the deviations, might affect the results. 
If the pseudonyms are not changed regularly, then privacy will be violated. 
More importantly, it is assumed that the nodes send correct locations. 
However, if the misbehaving nodes send their location in beacons, then a malicious node might send false location information at different sampling points
and might go undetected.

\section{Model, definitions and notations}
\label{sec:model_def_not}
In this section we present our model, definitions and notations which we use throughout the rest of the paper. 
\subsection{Our model and assumptions}
\label{subsec:model}
The network consists of a set $\mathcal{N}$ nodes of $|\mathcal{N}| = N$ nodes, a set of RSUs,   $\mathcal{R}$,  and 
a set of CAs, $\mathcal{C}$. 
Vehicles are denoted by $n_i$ (also called nodes), road side units (RSU) by $R_i$ and Certification Authorities (CA) $C_i$ and a Master Authority MA. 
We assume a \emph{one way traffic network, consisting of three lanes}.  

 The MA is headed by the government of a State or Province. Its authority is divided into several smaller regions each 
having a  local authority named CA. 
The CAs are government agencies that maintain records of vehicles and their owners, and issue
unique identities as license plates and secret credentials like pseudonyms, 
public/private keys and certificates. 
We use the authentication scheme ECMV \cite{WJS08} which completely suits our purpose.
It has an  hierarchical structure with several CAs. 
We assume that the CAs to be trustworthy and has
authority over all vehicles registered locally.

  We assume that RSUs are much more difficult to
compromise than the vehicles. Although RSUs are some-
times in isolated places, their hardware may present
some tampering proof capabilities, making it difficult
for a regular human being to compromise it. This is not
true in the case of vehicles, because its owner can drive
to an expert just to have the vehicle tampered.

Nodes misbehave by sending out false information, mainly out of selfish motives like getting faster and easier access of a road, 
getting credit card and other confidential information about fellow nodes. 
Depending on
their intention when sending a false information, nodes
can either be faulty (damaged), selfish or malicious.

Since most of the undesirable behaviors will be caused due to selfish motives, 
the classification of ``good" node and ``bad" node is not so important as the distinction between correct information and false information. 
We will be more
concerned with finding out if the message or alert signal
generated by a node is correct or false, than if the node
is ``good" or ``bad". 
This can be termed as \emph{data-centric misbehavior detection}. 
This is different from entity-centric MDS, where nodes are either classified as ``good" or ``bad".

A node can send several types of messages when on the road. 
We mainly deal with two types of messages. 

\begin{enumerate}

\item \emph{Alert messages} that ensure safety of vehicles on the road \cite{BKVHE06}. These include:

\begin{enumerate}
\item Emergency Electronic Brake lights (EEBL), 
\item Post Crash Notification (PCN), 
\item Road Hazard Condition Notification (RHCN), 
\item Road Feature Notification (RFN), 
\item Stopped/Slow Vehicle Advisor (SVA), 
\item Cooperative Collision Warning (CCW),
\item Cooperative Violation Warning (CVW),  
\item Congested Road Notification (CRN), 
\item Change of Lanes (CL),
\item Emergency Vehicle approaching (EVA). 
\end{enumerate}
\item \emph{Beacons} which specify the location of the vehicles. 
\end{enumerate}

The alert messages are important to send safety information, so that actions can be taken and accidents can be prevented. 
EEBL alerts that a vehicle is decelerating rapidly, so that the rear vehicles can prevent rear-end collisions. 
PCN alerts are sent by vehicles warning other vehicles of an accident which has already occurred. 
RHCN reports of road conditions like ``slippery road" or ``ice" or unwanted debris on the road. 
SVA alerts that a vehicle is moving slowly. 
RFN alerts of speed limits near schools and hospitals or a sudden bend or steep slope. 
CCW sends information about possible collisions that should be avoided. 
CVW warns vehicles about possible violations of traffic signals. 
CL notifies when a node is changing its lane.
These conditions are sent by nodes to nodes behind them. 
The alert EVA might also be sent by nodes to other nodes approaching from behind. 

Alerts can be either observed or self generated. 
For example, a vehicle might observe the road hazardous condition (PCN, RHCN) or ``school ahead" sign (RHN) or a slow moving vehicle (SVA). 
Alerts can also be generated by the node itself when it is decelerating rapidly (EEBL) or changing lanes (CL).

The CAs know the mapping between the pseudonym  used by each node
and the unique id relating the pseudonym to the node.
If there is evidence from the RSU and other nodes, that the node has misbehaved in a given situation, then this is noted and
a penalty is imposed in the form of a fine. 
This is very similar to the general practice of imposing fines.
The difference is that the decision is no longer taken by an authority, like police and speed cameras which might not be within reach
 when the event has happened, but taken by fellow nodes or RSUs. 
The idea of paying fines  will decrease misbehaving from regular vehicles.
We will assume, that a faulty vehicle will be condemned in a similar way. 
 Therefore, it is up to the owner to maintain  all vehicle's sensors at good conditions and always be aware 
of what messages are being sent out. 
Each alert has an associated  amount of fine, depending the impact of misbehavior.

We define \emph{freshness interval} as the time period during which a message is fresh.
This will vary for different types of alerts.  
We denote it by $F_T$. 
 
\subsection{Notations}
The Table 1 gives the notations that we follow throughout the rest of the paper. 
We use nodes and vehicles interchangeably.

\begin{table}
\begin{center}
\begin{tabular}{|c|c|}
\hline
Notation & Meaning \\
\hline
$N$ & Number of nodes in the network\\
$n_i$ & $i$th node\\
$R_i$ & $i$th RSU\\
$C_i$ & $i$th CA\\
$T$ & Type of alert\\
$F_T$ & Period of freshness \\
$p_{it}$ & Pseudonym of node $i$ at time $t$\\
$l_{it}$ & Location of $n_i$ at time $t$\\
$E_i$ & Event $i$ for which an alert is generated\\
$L_i$ & Location of the event $E_i$\\
$M_A$ & Alert message\\
$M_R$ & Replay message\\
$M_B$ & Beacon\\
$dist(l_{it},l_{jt})$ & Distance between $l_{it}$ and $l_{jt}$\\
\hline
\end{tabular}
\end{center}
\label{table:notations}
\caption{Table of notations}
\end{table}

\section{Proposed Misbehavior Detection scheme}
\label{sec:MDS}
We first give a sketch of our approach and then work out the details. 

\subsection{Sketch of our misbehavior detection system}
Suppose a node $n_j$ having a pseudonym $p_{jt}$ sends out an alert message $M_A$ at time $t$. 
Once a node $n_i$ receives some alert signal from a node $n_j$, it finds out from the alert message,  
the type of alert and the location of the event $E_x$ for which the alert was generated. 
For example, the alert might be ``There is road block in location X". 
In this case $E_x$ would be ``road block" and $L_x$ would be $X$. 
The type $T_x$ is RHCN. 

When $n_i$ later receives a beacon from node $n_j$ after an elapse of time $\Delta t$, it checks the current location of $n_j$, 
and checks if it can be a valid location for $n_j$. 
For example if node $n_j$ first sends a message ``There is a road block in location X" and after time $\Delta t$ it is close to location $X$, 
then it implies two contradictory statements and the alert message cannot be trusted. 
In this case  $n_j$ might send a false alert, $n_j$ sends to divert the traffic away from the location $X$. 

For this reason, each node maintains a list called the \emph{list of invalid events, (LIE)}, for short. 
LIE contains a list of events and corresponding invalid actions. 
For example, if there is an emergency breaking, then the distance between the old and new positions of $n_j$ cannot be more than 100 meters. 
In Table 2, we present event and invalid action pair. 

LIE will contain the information from first and third column of the table. 
The change of lanes can be known from the position information and interpreting it using GPS. 
$d$ is the safe distance.  For example if a car is driving at 80kmph when it observes the alert and 
then reduces its speed to 20kmph as a consequence of the alert, then it will travel less about 100 meters in the next two seconds. 
Thus the positions sent in the beacons will be less than $d=$100 meters apart. 

As soon as the node $n_i$ receives an alert message it transmits the message. 
Later, after time $\hat{t}$, if $n_i$ realizes that the message from node $n_j$ is incorrect, then it sends out the negation of the alert message 
already sent. 
Node $n_i$ checks for a certain time interval $\hat{t}$ and if it does not receive any beacon during this interval,
then it assumes that the node $n_j$ has changed its pseudonym. 
It also sends a message to the RSU, convicting $n_j$ of sending false alert message. 
RSU checks with its own observation and sends a message to the CA stating the misbehavior and the pseudonym of the node $n_j$. 
CA knows the mapping of the pseudonym with the original id and update its records against node $n_j$. 

\begin{table*}[!t]
\begin{center}
\caption{Events and invalid actions}
{\normalsize
\begin{tabular}{|c|c|c|}
\hline
Event & Expected action & Invalid Action\\
\hline
EEBL & Car must slow down & $D > d$ meters\\
PCN & Car stops/Changes Lane  & $D > d$ meters and No Lane Change\\
RHCN & Car stops/ Changes rout & $D > d $ meters and Same Route\\
RFN & Decrease speed & $D > d$ meters\\
SVA & Change lane/decrease speed & $D > d$ meters and Same Lane\\
CCW & Slow down & $D > d$ meters\\
CVW & Slow down & $D > d$ meters\\
CL & Lane change & Same Lane \\
EVA & Change Lane/slow down & $D > d$ meters and Same Lane as Vehicle\\
\hline
\end{tabular}
}
\end{center}
\begin{center} {\it Where $D \equiv dist(l_{jt_1},l_{jt_3})$. The first and third columns are used to build LIE.}\end{center}
\label{table:valid-invalid}
\end{table*}

\subsection{Misbehavior detection scheme in details}
There is a pool of pseudonym $\mathcal{P}$. 
A node $n_i$ is given a set of pseudonyms $P_i$ from this pool, such that 
$\bigcup_{i=1}^N P_i =\mathcal{P}$. 
Each node $n_i$ has an id $i$ and pseudonym $p_{it} \in P_i$, at time $t$. Let
 $\tau$ denote time. 
Each vehicle has an on board unit (OBU), 
which is loaded with a public/private key pair, corresponding to each pseudonym,  by a CA.

We use the certificate management scheme, ECMV of Wasef et al \cite{WJS08}. 
There is a master authority (MA) and several CAs. 
The MA generates public/private key pairs and two secret certificate signing keys, for each CA. 
The MA also generates public keys for verifying the certificates of RSUs and nodes. 
Each CA uses the certificate signing keys to sign a certificate set of each RSU. 
The certificates in a set are shared among all the RSUs in the set. 
The other secret key is used to generate  a partial signing key for each RSU. 
Each RSU uses this partial signing key to generate certificates for each node within its range. 
Public keys can be used by CAs, RSUs or nodes to verify the certificates of RSUs and nodes. 
ECMV has certificate update algorithm which suits our purpose fully. 
For more details on the ECMV scheme one can refer to \cite{WJS08}.

An event $E_x$ can be for example like:
\begin{enumerate}
\item Emergency breaking,
\item Observation of unwanted debris on road or hazardous road conditions like ``ice", ``slippery road" etc,
\item Observation of ``Drive slow" sign in areas like school or hospitals or steep slope,
\item Crash Notification,
\item Approaching emergency vehicles.
\end{enumerate}
A list of emergency alerts has been discussed in \ref{subsec:model}. 
We denote the set of alerts by $\mathcal{T}$. 
The set of locations is given by $\mathcal{L}$.  
$\mathcal{M}$ denotes the message space. 

An  \emph{alert message}, denoted by $M_A \in \mathcal{M}$ is a five tuple 
\begin{center}
$M_A = (p_{it},T,L_j,t,l_{it})$, 
\end{center}
where,
$p_{it}\in P_i$ is the pseudonym of the node $n_i$ who generated the alert at time $t \in \tau$, \\
$T \in \mathcal{T}$ is the type of alert, which can be one of the alerts which have already discussed, \\
$L_j \in \mathcal{L}$ is the location of the event $E_j$ for which the alert was generated, \\
$t \in \tau$ is the time at which the alert message had been sent, \\
$l_{it} \in \mathcal{L}$ is the location of the node $n_i$ which generated the alert at time $t$.

A node that receives an alert message from a neighboring node, relays it to other nodes and RSUs in its vicinity.

A \emph{relay alert}, denoted by $M_R$ is a  tuple, 
\begin{center}
$M_R = (p_{it},t,M_A)$,
\end{center}
where, 
$p_{it}\in P_i$ is the pseudonym of the node $n_i$ who sends the relay alert, 
$t \in \tau$ is the time at which $M_R$ was sent and 
$M_A \in \mathcal{M}$ is the alert message that it is relaying.

A \emph{beacon} sent by a node is denoted by $B$ and is a three tuple
\begin{center}
$M_B = (p_{it}, t, l_{it})$,
\end{center}
where, 
$p_i \in P_i$ is the pseudonym of the node, 
$t \in \tau $ is the time at which the beacon was sent, and
$l_{it} \in \mathcal{L}$ is the location of the node.

Suppose a node $n_i$, having a pseudonym $p_{it_2}$ at time $t_2$ receives an alert message $M_A$, 
from a node $n_j$ (with pseudonym $p_{jt_1}$), it first checks if it has a valid signature. 
This can be done by ECMV \cite{WJS08}. 
If the node $p_{jt_1}$ has a valid signature, then $n_i$ notes the time $t_1$ from the message $M_A = (p_{jt_1},T,L_x,t_1,l_{jt_1})$. 
$L_x$ is the location of the event for which the alert was generated. In some cases, for example
lane change information it is possible that $L_x = l_{jt_1}$. 

We define a threshold time $F_T$ after which a message becomes stale. 
$F_T$ is also known as \emph{period of freshness}. 
If $t_2-t_1 > F_T$, then it means that $n_j$ had sent it long back and has  become stale. 
So $n_i$ discards the message $M_A$. 
If the message is fresh then the position of the event $L_x$ and the location of node $l_{jt_1}$ is noted. 
If the positions are contradictory, then no action is taken for the alert and the message is discarded. 

The positions are contradictory, if the order of location is anything other than 
$n_i-n_j-E_x$ or $E_x-n_j-n_i$. 
The first condition arises when the event has occurred in front of $n_j$ and $n_j$ is sending a message to the node $n_i$ behind it. 
The second condition arises when the event has occurred  behind $n_j$ and $n_j$ is sending a message to the node $n_i$, which is in front of it. 
The first condition arises when there is an accident at $E$ and emergency breaking alert is raised by $n_j$, 
or there is road hazard like water or ice on road. 
The second order arises if there is an emergency vehicle approaching from behind. 
So node $n_j$ reports to $n_i$ (who is in front of it) to make space for the vehicle approaching from behind. 
 
If the positions are correct, then the node $n_i$ considers the alert.
We will see in the next section, how to detect incorrect location information. 
It checks the alert type and prepares to take action against it. 
The node $n_i$ might receive more that one alert messages from different nodes.
We do not, however make a decision on the validity of the alert based on the number of vehicles 
that report the alert, because we do not rely on thresholds.
This is where our scheme differs from other VANET MDS. 
For this reason, Sybil attack is not effective against our scheme. 

After receiving an alert message, the node $n_i$ waits for beacons from $n_j$ for a  time period of $\hat{t}$. 
It verifies the position of the node $n_j$ from all the beacons it receives during this time period. 
When the node $n_i$  receives beacon message from the nodes $n_j$, 
it checks the position $L_x$ in the alert message
and the position $l_{jt_3}$ of the beacon message $M_B = (p_{jt_3}, t_3, l_{jt_3})$. 
It checks the LIE to see if the alert type contradicts with the position. 
If it does, then $n_i$ sends out an alert message which is the negation of the previous message. 
It also reports the nearest RSU and convicts node $n_j$ for sending false message. 
If node $n_i$ does not receive any beacon from $n_j$ in time $\hat{t}$ after receiving an alert message from $n_j$, 
then it assumes that the pseudonym has changed.
Changing pseudonym within a time $\hat{t}$ is considered to be a misbehavior and so $n_i$ reports the RSU that $n_j$ is misbehaving. 

The RSU upon receiving such conviction messages, compares with its own observation and reports
to the CA the pseudonym of the misbehaving node along with the reason for accusing it.  
Only the CA can match the pseudonym with the original identity of the node. 
The CA then  issues negative points to the node, which has to pay it as a fine, depending on the number of negative points received. 
Our assumptions are based on the fact that, any misbehaving node does so, mainly due to  selfish reasons and is most likely 
not to send false message all the time, but only when needed. 
A large number of misbehaviors can be interpreted as a malicious motive and such nodes can be revoked off their certificates and
other secret credentials using the revocation scheme in PASS \cite{SLLSS10}.

\subsection{How to detect incorrect location information}
In the previous section, we assumed that the location information send in the alert message or in the beacons is correct.
However, a clever malicious node will also send incorrect location information, along with the false alert message. 
In this section we see, how to detect incorrect location information. 
Studder et al \cite{SLP07} have presented how to detect nodes moving in a straight line  
and transmitting false location information.
The decision to convict a node depends on the number of votes  cast against it. 
If a node is surrounded by many corrupt nodes, then a node cannot be convicted. 
The authors also show that if the first two nodes in the straight line (convoy) send false messages, 
then they cannot be detected. 
In our scheme it is very likely that the first node transmits a false alert or beacon with wrong position information. 
So we cannot use the limited incorrect location detection approach of Studder et al. 

Suppose a node $n_j$ sends a beacon at time $t_1$, then suppose $n_i$ (in communication range of $n_j$) receives the message at time $t_2$. 
Then $t_{2}$ is given by 
\begin{equation}
t_{2} = t_{1} + \frac{dist(l_{it_2},l_{jt_1})}{c}
\label{eq:location}
\end{equation}
where $c$ is the speed of light. 
Suppose, node $n_j$ wants to fake its location as $l'_{jt_1}$, so it sends a beacon with the information
$(p_{jt_1},t_1,l'_{jt_1})$. Node $n_i$ receives it at time $t_2$. 
Node $n_i$ finds out that $n_j$ is lying because Eq(\ref{eq:location}) does not hold. 
To convince $n_i$,  that it is not lying, node $n_j$ should also change the time stamp when the beacon is sent. 
The time $t_1'$ at which he must send the
message, so that $n_j$ receives it at $t_2$, is given by
\begin{equation}
t_2 = t_1' + \frac{dist(l_{it_2}, l'_{jt_1})}{c}
\end{equation}
So, 
\begin{equation}
t_1' = t_{1} +  \frac{dist(l_{it_2},l_{jt_1})}{c} - \frac{dist(l_{it_2}, l'_{jt_1})}{c} 
\end{equation}
So node $n_j$ sends a beacon $(p_{jt_1}, t_1', l'_{jt_1})$ to convince $n_i$ that it is sending the correct message. 
However, since node $n_j$ does not know the distance  between itself and the node $n_i$ accurately,   
it cannot accurately calculate $t_1'$. 
So, when node $n_i$ observes the time stamp $t_1$ and the false location $l'_{jt_1}$, then any node can 
calculate the expected position and verify it according to equation (\ref{eq:location}).

\begin{figure}[htb]
\begin{centering}
\includegraphics[width=2.5in]{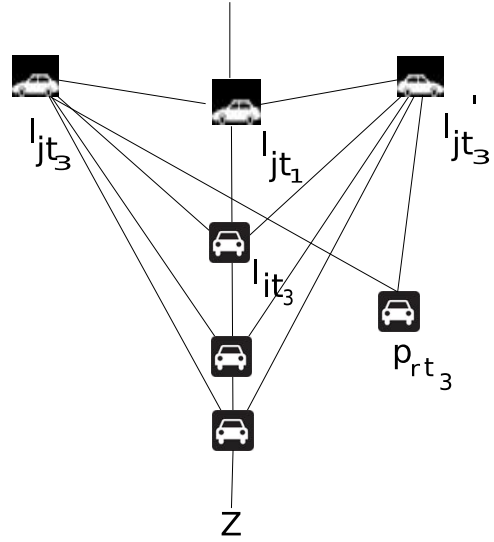}
\caption{
Exception when many cars are equidistant from the true and correct positions of $n_j$
}
\label{fig:exception}
\end{centering}
\end{figure}

There is an exception to the above condition. 
We refer to the Fig \ref{fig:exception}. 
In case the node sends a location information $l_{jt_3}'$ at time $t_3$, when its current location is $l_{jt_3}$, 
then $n_i$ will not be able to make out that node $n_j$ is lying because it is equidistant from both
the locations $l_{jt_3}'$ and $l_{jt_3}$. 
Similar is the case for all vehicles which lie on the line $Z$. 
However, if there is another node $n_r$ (with pseudonym $p_{rt_3}$, as shown),  then it will be able to make out that $n_j$ is lying about its position. 
If the RSU is not on $Z$, then it is able to understand that $n_j$ is lying and  it will take action against it. 
In the rare case if the RSU is also on the line $Z$, it cannot predict correctly based on one node's correct observation.

We will include this observation in the MDS of previous section. 
On receiving the message $M_A$, a node $n_j$ first checks the authenticity of the message using ECMV \cite{WJS08}. 
If the message is not stale ($t_2-t_1 <F_T$, Step 5) and the order of the vehicles is either $n_i-n_j-E_x$ or $E_x-n_j-n_i$ (checked by the condition 
$dist(l_{jt_1},l_{it_2})< dist(l_{{it_2},L_x})$), then node $n_i$ waits for beacons  from $n_j$. 
It checks this for time $\hat{t}$. 
Suppose  $n_i$ receives a beacon message $M_B = (p_{jt_3}, t_3,l_{jt_3})$ from $n_j$ at time $t_3$. 
$n_i$ first checks the validity of the location using Eq(\ref{eq:location}). 
It then looks up in LIE for event $E_x$. 
If the action is invalid, then it reports misbehavior and broadcast the negation of the message. 
If action is correct, then it broadcasts the message $M_B$. 
The variable count, keeps track of the number of beacons received during the interval $\hat{t}$. 
If no beacon is received, then it implies that the node $n_j$ has changed its pseudonym and this is reported and misbehavior reported.

We present the MDS algorithm below. 

\begin{algorithm}
\caption{MDS algorithm operated by a node $n_i$ }
\label{algo:MDS}
\textbf{Input}: Alert message $M_A$, beacons $M_B$, Table LIE\\
\textbf{Output}: ''Valid Alert" or ''Invalid Alert"
\begin{algorithmic}[1]
\STATE Tag =1, count = 0 
\STATE  Node $n_i$ receives  $M_A = (p_{jt_1}, T, L_x, t_1, l_{jt_1})$ at time $t_2$
\STATE Check authenticity of $M_A$ using ECMV
\IF{$M_A$ is authentic and Eq(\ref{eq:location}) holds} 
\IF{($t_1< t_2$ and  $t_2 - t_1 < F_T$) and ($dist(l_{jt_1},l_{it_2})< dist(l_{{it_2},L_x})$)}
\WHILE{$ t < t_2 + \hat{t}$}
\WHILE{$n_i$ receives beacon from $n_j$}
\STATE  Node $n_i$ receives  $M_B = (p_{jt_3}, t_3, l_{jt_3})$ at time $t_4$
\STATE  count = count +1
\IF{$t_3$ and $t_4$ satisfy equation(\ref{eq:location})}
\STATE Look up in LIE for type $T$ of event $E_x$
\STATE $r = $ check\_action\_function(LIE,$T$, $t_1$, $t_3$, $l_{jt_1}$, $l_{jt_3}$)
\STATE Tag = 0
\IF{$r=1$}
\STATE Broadcast $M_R = (p_{it_5}, t_5, M_A)$
\STATE Take action against $E_x$
\STATE Print ``Valid Alert"
\ELSE 
\STATE $M_A'$ = Message\_negation($M_A$) \COMMENT{creates the negation of the message}
\STATE Broadcast $M_R = (p_{it_5}, t_5, M_A')$
\STATE Tag = 1
\ENDIF
\ELSE 
\STATE Tag = 1
\ENDIF
\ENDWHILE
\ENDWHILE
\IF{Count =0}
\STATE ``Pseudonym Change"
\STATE Tag = 1
\ENDIF
\ELSE 
\STATE Tag = 1
\ENDIF
\ELSE 
\STATE Tag = 1
\ENDIF
\IF{Tag = 1}
\STATE report\_misbehavior($p_{it_5},M_A$) 
\STATE Discard $M_A$
\STATE Print ``Invalid Alert"
\ENDIF
\end{algorithmic}
\end{algorithm}

The procedure check\_action\_function takes the inputs LIE, $T$, $t_1$, $t_3$, $l_{jt_1}$, $l_{jt_3}$ and outputs
$0$ or $1$ depending on bad or good behavior. 
First it finds out the event $E_x$ that cause the alert. 
Then it looks up in the table LIE to check if the conditions corresponding to $E_x$ if the conditions hold. 
If the conditions match $r$ is set to 0, meaning that there is a misbehavior. 
If the conditions in the table LIE do not match, then $r =1$. 

The function report\_misbehavior takes as input the pseudonym of the reporting node and the false alert message and send it to the RSU. 
Message\_negation creates the negation of the message. 
For example if the alert was ``There is ice on rout X", then the negation of the message is 
``There is no ice on rout X". 
If node $n_i$ retransmits a false message, then it will be found in either of the two ways: 
\begin{enumerate}
\item By node $n_i$ (if within communication range) by
observing that $M_R$ does not contain the message $M_A$ that it had sent. 
\item By other nodes which receive $M_R$ and subsequent location information from beacons of $n_i$.
They will find out if node $n_i$ is malicious by an algorithm similar to Algorithm \ref{algo:MDS}. 
\end{enumerate}

However it is possible that node $n_i$ is falsely accusing node $n_j$. 
In this case the RSU will reject the accusation and convict $n_i$. 
The other cars will be warned that $n_i$ is lying. 

Since the results are also verified by the RSU before sending to the CA, even if the observing vehicles are misbehaving, the 
misbehavior can be detected. 
The above approach does not require any voting or majority, so a Sybil attack does not have any effect in 
misbehavior detection. 

\begin{table*}
\begin{center}
{\normalsize
\begin{tabular}{|c|c|c|c|}
\hline
Scheme & Location Privacy & False Location Info. & Immune to Sybil Attacks\\
\hline
LEAVE \cite{RPAJH07} & Yes & Yes & No \\
Golle et al \cite{GGS04} & No & Yes & No\\
Zhou et al \cite{ZCNC07} & No & No & Yes\\
Ghosh et al \cite{GVGKM10} & No & No & No \\
Ours & Yes & Yes & Yes\\
\hline
\end{tabular}
}
\end{center}
\caption{Comparison with other misbehavior detection schemes}
\label{table:other_MDS}
\end{table*}

\subsection{How to deal with compromised RSUs}
RSUs are prone to be compromised.
However, compromising RSU is much difficult that compromising nodes.  
Compromised RSUs can either transmit false messages or convict benign nodes.
A RSU which transmits false messages can be noticed by other nodes, which reports it to the next RSU. 
If it receives a large number  of such reports over a long time, then the RSU is considered to be compromised
If benign nodes are convicted, they are imposed fines by the CA. 
If the message sent by a node has been modified by the RSU, then  the convicted node can prove it using its signature. 
We note here that in ECMV certificates are not created by the RSUs. So RSUs cannot fake the signature on the messages, sent by the nodes.
The nodes can then prove their message authenticity, using the signatures on the messages. 

Once it is known that an RSU has been compromised, its certificate is revoked using the techniques in ECMV scheme. 
We note that the RSUs much fewer in number, compared to the nodes. So
broadcasting a CRL for RSUs will not be expensive. 
If the number of misbehaviors observed for a node is very high, then it is assumed that the node is 
malicious and removed from the network altogether.

\section{Performance analysis and comparison}
We compare our scheme with existing MDS schemes in Table 3. 
Our scheme has all the desirable 
properties like location  privacy, ability to detect false location information and immunity against Sybil attacks.

In our scheme we have used ECMV \cite{WJS08} for authentication which fits into the hierarchical
structure of our network. 
The transmission delay is only $6.47ms$ (as stated in \cite{WJS08}). 
The time taken for certificate verification is
$14.7$ms and for signature verification is $5.1ms$. 
($3T_{par} + 2T_{mul}$ and $2T_{par} + T_{mul}$ where $T_{par} = 4.5ms$ and $T_{mul}= 0.6$ms). 

\begin{figure}[htb]
\begin{centering}
\includegraphics[width=3.75in]{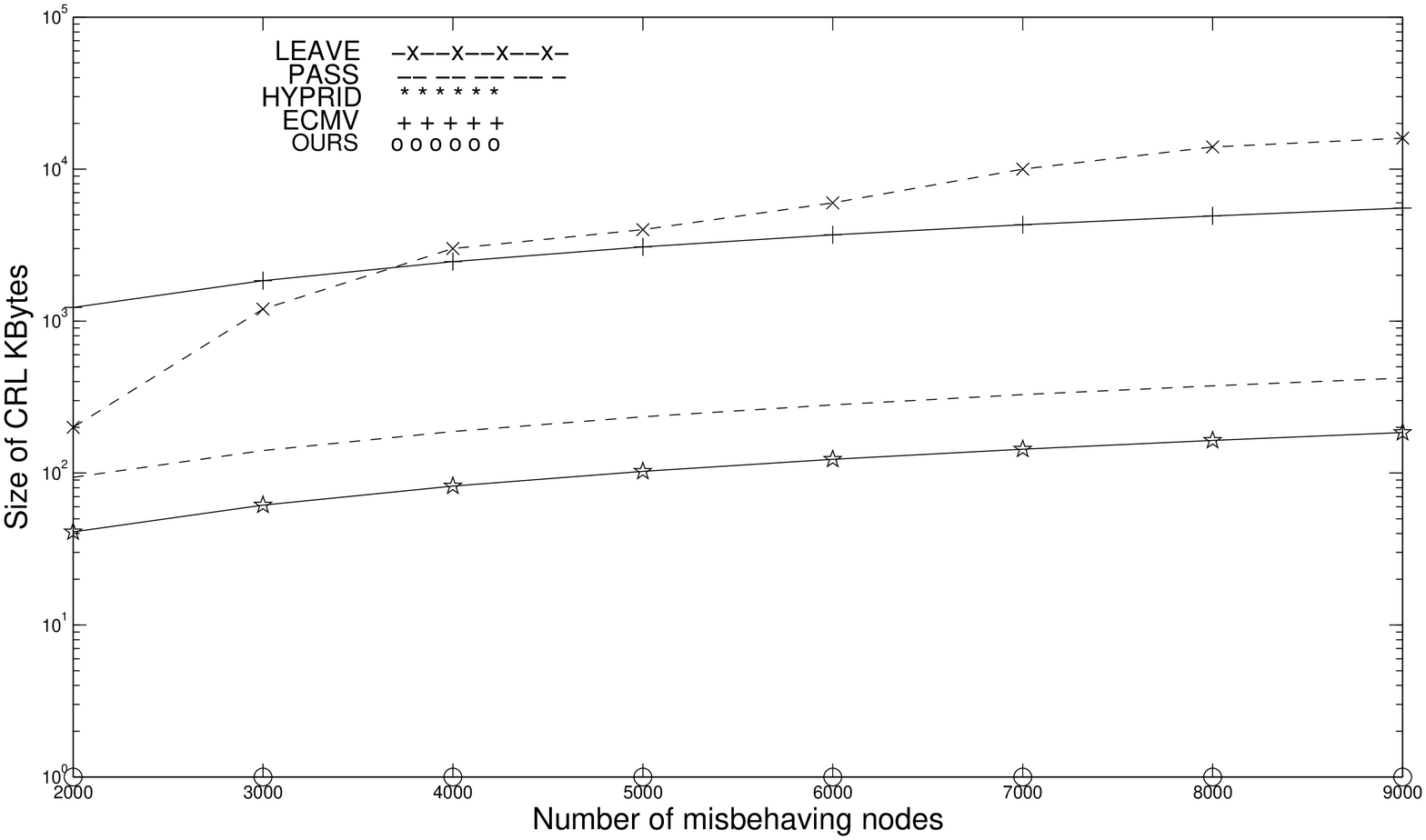}
\caption{
Comparison of communication overhead required for to send the CRL}
\label{fig:compare}
\end{centering}
\end{figure}

According to our algorithm, there is no extra communication overhead for revocation list, because only the 
identity of the misbehaving node is sent to the nearest RSU. 
The other nodes need not check any CRL list before then send their message. 
The message transmitted is simply the negation of the transmitted message which involves the same cost. 
Since CRL is not needed in our scheme, communication overhead is greatly reduced as compared to other schemes. 
In Fig \ref{fig:compare}, we show the communication overhead incurred by different schemes. 
We consider the Hybrid scheme by Calandriello et al \cite{CPHL07}, ECMV scheme \cite{WJS08}, PASS \cite{SLLSS10} and LEAVE \cite{RPAJH07}. 
We see that LEAVE incurs a high communication overhead, compared with \cite{CPHL07}, \cite{WJS08} and \cite{SLLSS10}. 

\section{Limitations and countermeasures: future course of action}
\label{sec:limitations}
In this section we discuss certain limitations our scheme and provide possible solutions to the problems.
\subsection{Limitations}
Our scheme is based on calculating distance between nodes, in order to detect false location information. 
We have already shown one condition (Fig \ref{fig:exception}), where a node $n_j$ will not be detected to be misbehaving.
Another situation where our scheme will fail to give correct results is when a node takes a U-turn. 
In such a situation, it might seem that a node is giving false location information, 
because its location is far behind the expected location.
It might seem that the vehicle has moved backwards, which means its location information is faulty.  
The other situation is when a node is moving on a flyover with loops, 
the actual distance might be quite different from the Euclidean distance $dist(u,v)$. 
We leave it as an open problem.

Another limitation is when nodes are moving in a group. 
One node might aid the other by sending a wrong alert information. 
Suppose two nodes $n_i$ and $n_j$ are moving together as a group. 
Node $n_i$ should take a right turn and node $n_j$ is supposed to go left. 
Node $n_i$ takes the right turn as per its requirement. 
Node $n_j$ can then aid node $n_i$ by sending an alert like ``hazardous condition on right road" and take the left one as desired. 
Nodes behind $n_j$ notice the alert and also the subsequent beacon messages and conclude that the alert sent by $n_j$ is correct, following our algorithm. 
Misbehavior of node $n_j$ will go undetected. 
They might then take the straight road and benefit node $n_i$. 
Such issues are left as future work. 

Another limitation is that, since each node retransmits the alert message to the others, 
the same alert message will be retransmitted by several nodes, using more bandwidth than necessary. 
How to efficiently handle this problem is left as a future work. 

\subsection{Incentivizing nodes}
We assume that nodes which sense or receive alerts, immediately transmit them to other nodes. 
However transmitting such alerts leads to power consumption.
So, it is natural to ask why would a node be motivated to send out such alerts, when its only goal is to save itself. 
One way to overcome this problem is to give incentives to nodes which co-operate in 
transmitting useful messages. 
So that, every transmitted message $M_A$ and $M_R$ is noted. 
The node $p_{jt}$ is given an incentive, in terms of positive points is it sends out correct information and
given negative points, if it sends out wrong information. 
The points can be suitably adjusted as providing free service for points above a certain threshold. 

\subsection{Change of direction}
We have assumed that nodes move in the same direction. However, a node moving in the opposite direction can send  out
false alert message. Such nodes might not do so for selfish reason, but have malicious intentions. 
We leave this problem open for future research. 

\section{Conclusion}
\label{sec:conclusion}
We have discussed the limitations of existing misbehavior detection schemes in VANETs and proposed a new scheme.
We introduce the concept of data-centric MDS, where we are more interested in finding out
false information than classifying nodes as ``good" and ``bad". 
The main reason for this is that nodes misbehave mainly because of selfish reasons and 
need not be classified as ``good" or ``bad" nodes. 
Our scheme provides location privacy, by the use of pseudonyms. 
Any node can detect false alert information, by observing the location of the node after issuing an alert. 
There is no need of voting and majority decisions. 
This makes our scheme resilient to Sybil attacks. 
So we do not revoke nodes completely, but impose fines depending false message sent out. 
We point our some limitations of our scheme and discuss several directions for future work.

\bibliographystyle{plain}

\end{document}